\documentclass[journal=jacsat,manuscript=communication,etalmode=truncate,keywords=true]{achemso}
\setkeys{acs}{articletitle=true,maxauthors=10,etalmode=truncate,keywords=true}
\usepackage{amsmath,amssymb,units,times}
\usepackage{amsfonts}
\usepackage{graphicx,MnSymbol}
\usepackage{helvet,wasysym}
\usepackage[bookmarks=true,colorlinks=true,urlcolor=blue,linkcolor=blue,citecolor=blue]{hyperref} 
\usepackage[usenames,dvipsnames]{color}
\usepackage{colortbl,cuted}
\usepackage{xr}

\definecolor{JPCTCGreen}{RGB}{11,122,64}
\definecolor{abstractcolor}{RGB}{255,243,201}
\makeatletter\newenvironment{abstractbox}{%
   \begin{lrbox}{\@tempboxa}\begin{minipage}{0.988\textwidth}}{\end{minipage}\end{lrbox}%
   \colorbox{abstractcolor}{\usebox{\@tempboxa}}
}\makeatother

\renewcommand*\authorfont{\flushleft}
\renewcommand*\affilfont{\mdseries\flushleft\textrm}
\renewcommand*\titlesize{\Large}
\renewcommand*\titlefont{\flushleft\sf\bfseries}
\def\refname{\sffamily\color{JPCTCGreen}\Large$\blacksquare$\normalsize\ REFERENCES}
\usepackage{titlesec}
\titleformat{\section}{\bfseries\sffamily\color{JPCTCGreen}}{}{0pt}{\large$\blacksquare$\normalsize~}
\titleformat{\subsection}[runin]{\bfseries\sffamily\normalsize}{\indent\thesubsection.~}{0pt}{}[.]
\titlespacing{\subsection}{0pt}{0pt}{*1}
\titleformat{\subsubsection}[runin]{\bfseries\sffamily\normalsize}{\indent}{0pt}{}[.]
\titlespacing{\subsubsection}{0pt}{0pt}{*1}

\newcommand{\old}[1]{}

\title{Quasiparticle level alignment for photocatalytic interfaces}

\author{Annapaoala Migani}
\email{amigani@cin2.es}
\affiliation[ICN2]{\footnotemark[2]{\ } ICN2 - Institut Catal\`{a} de Nanoci\`{e}ncia i Nanotecnologia and CSIC - Consejo Superior de Investigaciones Cientificas, ICN2 Building, Campus UAB, E-08193 Bellaterra (Barcelona), Spain}
\alsoaffiliation[UPV/EHU]{\newline\footnotemark[3]{\ } Nano-Bio Spectroscopy Group and ETSF Scientific Development Center, Departamento de F{\'{\i}}sica de Materiales, Centro de F{\'{\i}}sica de Materiales CSIC-UPV/EHU-MPC and DIPC, Universidad del Pa{\'{\i}}s Vasco UPV/EHU, E-20018 San Sebasti\'{a}n, Spain}
\author{Duncan J. Mowbray}
\email{duncan.mowbray@gmail.com}
\affiliation[UPV/EHU]{\newline\footnotemark[3]{\ } Nano-Bio Spectroscopy Group and ETSF Scientific Development Center, Departamento de F{\'{\i}}sica de Materiales, Centro de F{\'{\i}}sica de Materiales CSIC-UPV/EHU-MPC and DIPC, Universidad del Pa{\'{\i}}s Vasco UPV/EHU, E-20018 San Sebasti\'{a}n, Spain}
\author{Jin Zhao}
\affiliation[USTC]{\newline\footnotemark[5]{\ } Department of Physics and ICQD/HFNL, University of Science and Technology of China, Hefei, Anhui 230026, China}
\alsoaffiliation[SICQI]{\newline\footnotemark[4]{\ } Synergetic Innovation Center of Quantum Information \& Quantum Physics, University of Science and Technology of China, Hefei, Anhui 230026, China}
\author{Hrvoje Petek}
\affiliation[UP]{\newline\footnotemark[6]{\ } Department of Physics and Astronomy, University of Pittsburgh, Pittsburgh, Pennsylvania 15260, USA}
\author{Angel Rubio}
\email{arubio@ehu.es}
\affiliation[UPV/EHU]{\newline\footnotemark[3]{\ } Nano-Bio Spectroscopy Group and ETSF Scientific Development Center, Departamento de F{\'{\i}}sica de Materiales, Centro de F{\'{\i}}sica de Materiales CSIC-UPV/EHU-MPC and DIPC, Universidad del Pa{\'{\i}}s Vasco UPV/EHU, E-20018 San Sebasti\'{a}n, Spain}

\begin{document}

\maketitle

\begin{strip}
\vspace{-1.cm}

\noindent{\color{JPCTCGreen}{\rule{\textwidth}{0.5pt}}}
\begin{abstractbox}
\begin{tabular*}{17cm}{b{8.05cm}r}
\noindent\textbf{\color{JPCTCGreen}{ABSTRACT:}}
Electronic level alignment at the interface between an adsorbed molecular layer and a semiconducting substrate determines the activity and efficiency of many photocatalytic materials.  Standard density functional theory (DFT) based methods have proven unable to provide a quantitative description of this level alignment. This requires a proper treatment of the anisotropic screening, necessitating the use of quasiparticle (QP) techniques. However, the computational complexity of QP algorithms has meant a quantitative description of interfacial levels has remained elusive.  We provide a systematic study of  a prototypical interface, bare and methanol covered rutile TiO$_2$(110) surfaces, to determine the type of many-body theory required to obtain an accurate description of the level alignment.   This is accomplished via a direct comparison with metastable impact electron spectroscopy (MIES), ultraviolet photoelectron spectroscopy (UPS) and two-photon photoemission (2PP) spectra.  We consider GGA DFT, hybrid DFT and
&\includegraphics[width=8.5cm]{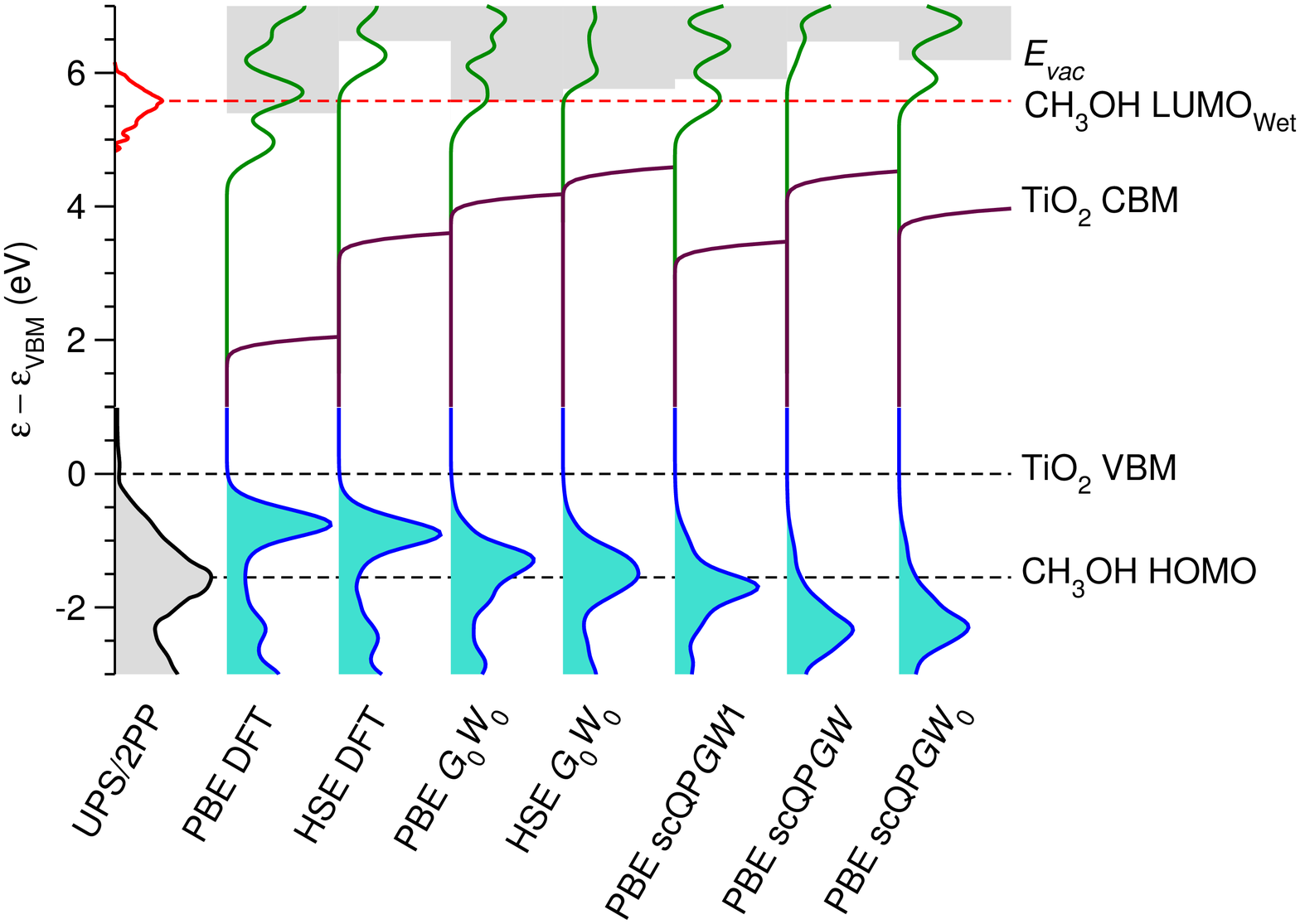}\\
\multicolumn{2}{p{17cm}}{
$G_0W_0$, $\mathrm{scQP}GW1$, $\mathrm{scQP}GW_0$, and $\mathrm{scQP}GW$ QP calculations.  Our results demonstrate that $G_0W_0$, or our recently introduced $\mathrm{scQP}GW1$ approach, are required to obtain the correct alignment of both highest occupied and lowest unoccupied interfacial molecular levels (HOMO/LUMO). These calculations set a new standard in the interpretation of electronic structure probe experiments of complex organic molecule--semiconductor interfaces.
}
\end{tabular*}
\end{abstractbox}
\noindent{\color{JPCTCGreen}{\rule{\textwidth}{0.5pt}}}
\end{strip}

\section{INTRODUCTION}

Describing the level alignment in photocatalytic materials is a fundamental first step in their development into technology. This requires a quantitative description of the levels associated with the  interface between a solid semiconductor and an adsorbed organic  molecule. The alignment of the frontier highest occupied and lowest unoccupied molecular orbitals (HOMO/LUMO) with the valence band maximum (VBM) and conduction band minimum (CBM) of the semiconducting substrate controls the interfacial electron transfer \cite{HendersonSurfSciRep,YatesChemRev}. Such electron transfer is necessary for a system to be photocatalytically active.  

For a theoretical technique to be robust for describing the interface, ideally, it would reproduce the electronic and optical band gaps, along with the optical spectra of the bulk material.  Further, the alignment of the VBM and CBM of the bare surface would be reproduced.  These requirements could then be used to justify the use of a theoretical technique for an interfacial system.  Only by reproducing the measured level alignment may we provide a robust theoretical interpretation.

The accurate description of level alignment requires techniques which incorporate the spatial dependence of the electron-electron correlation.  This is especially important for the interface, as it includes both the vacuum and the bulk substrate.  For unoccupied levels located in the vacuum above the surface there is little to no electronic screening.  This is already described by the bare Hartree term within standard density functional theory (DFT).  For unoccupied levels located within a semiconducting substrate, electronic screening can be quite significant.  This can be described by a constant screening of the Hartree-Fock exact-exchange term, as done via the fraction of exact-exchange included within hybrid DFT exchange and correlation (xc)-functionals \cite{Marques}. However, for unoccupied interfacial levels, with weight in both the vacuum and substrate, the screening is intermediate and spatially heterogeneous.  This anisotropy of the electron-electron correlation can be described using many-body quasiparticle (QP) techniques \cite{GW,AngelGWReview,Galli}.  This anisotropic screening is even more important for the levels of a molecular monolayer (ML) on a semiconducting substrate, e.g. image states \cite{ImageStatesLiFMgO} or wet electron levels \cite{OurJACS}.

Despite its importance, computations of  interfacial levels employing QP techniques are scarce \cite{OurJACS,Migani2014, GiustinoPRL2012, NaClCorrelation, RenormalizationLouie, JuanmaRenormalization1}. This  is mainly due to the difficulty in carrying out QP  calculations on atomistic models with hundreds of atoms, as they are  computationally prohibitive. Here, we apply  QP techniques to the accurate computation of the interface  between rutile TiO$_2$(110) and methanol.  Methanol is chosen for its important applications in photocatalysis (direct photocatalytic dissociation \cite{MethanolSplitting2010,MethanolSplitting2013}, hydrogen formation \cite{HydrogenfromMethanolPhotocatalysis}, photo-oxidation to formaldehyde \cite{MethanolPhotocatalysis} or methyl formate \cite{Friend2012, Weixin}) and photoelectrocatalysis (as a sacrificial agent \cite{Kawai1980} in oxidative dehydrogenation of water \cite{FujishimaNature,FujishimaReview}).  Viewed from a theoretical perspective, methanol on TiO$_{\text{2}}$(110) represents one of the  ``simplest'' and computationally feasible systems for applying QP $GW$ calculations to an entire interface \cite{OurJACS,Migani2014}.

Many studies have probed the electronic structure and photocatalytic activity of methanol on the single crystal rutile TiO$_{\text{2}}$(110) surface under ultra-high vacuum (UHV) conditions.  Experimentalists have employed a full arsenal of techniques, such as ultraviolet, X-ray, and two photon photoemission spectroscopy (UPS, XPS, and 2PP) \cite{Onishi198833,Weixin,PetekScienceMethanol,MethanolSplitting2010,PetekScienceH2O}, scanning tunnelling microscopy \cite{MethanolSplitting2010,HendersonReview,Henderson2012} (STM), and mass spectrometric analysis of reaction products \cite{Henderson2011,Friend2012,MethanolPhotocatalysis}.  A proper interpretation of these results requires a similar arsenal of robust theoretical techniques for their explanation.

\begin{figure}[!t]
\includegraphics[width=0.515\columnwidth]{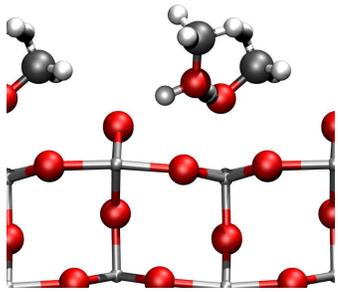}
\caption{Schematic of the most stable geometry for a CH$_3$OH ML on a TiO$_2$(110) surface from ref.~\citenum{Jin}.  H, C, O, and Ti atoms are represented by white, grey, red, and silver balls, respectively. Adapted with permission from ref.~\citenum{OurJACS}. Copyright 2013 American Chemical Society.}\label{schematic}
{\color{JPCTCGreen}{\rule{\columnwidth}{1.0pt}}}
\end{figure}

In this article we apply many-body QP techniques ($G_0W_0$ \cite{KresseG0W0}, $\mathrm{scQP}GW1$ \cite{OurJACS}, $\mathrm{scQP}GW$ \cite{SchilfgaardeQSGW,QSGW,KressescGW}, or $\mathrm{scQP}GW_0$ \cite{KresseGW}) to bulk (rutile TiO$_2$), surface (TiO$_2$(110)), and interfacial (CH$_3$OH ML on TiO$_2$(110) shown in Figure~\ref{schematic}) systems.  These are based on DFT calculations using either a local density approximation (LDA) \cite{LDA}, generalized gradient approximation (PBE) \cite{PBE} or hybrid (HSE) \cite{HSE} xc-functional. 
For the interface, the HSE $G_0W_0$ results of the present study are compared with those we reported previously in refs.~\citenum{OurJACS} and \citenum{OurJACSSI}.  Through this detailed comparison, our aim is to demonstrate not only which techniques provide the best description of the level alignment, but also why.  To accomplish this, we directly compare with: (1) the dielectric function probed by optical reflectivity \cite{OpticalTiO2} and ellipsometry \cite{SPIE} experiments for the bulk; (2) the work function probed by metastable impact electron spectroscopy (MIES) \cite{C0CP02835E} for the surface; and (3) the occupied molecular levels probed by UPS \cite{Onishi198833} experiments and the unoccupied molecular levels probed by 2PP \cite{Onda200532,PetekScienceMethanol,MethanolSplitting2010,Methanol2PPYang} experiments for the interface.  

It is the alignment of these molecular levels with the VBM and CBM which controls the photocatalytic activity of the interface.   However, for titania it is rather difficult for a single technique to describe the electronic and optical band gaps, optical spectra, and surface work function (i.e.\ VBM and CBM relative to the vacuum level) simultaneously.  We will show that a mixture of several techniques is necessary to obtain a complete description of the level alignment of titania bulk, surface, and interfacial systems.

\section{METHODOLOGY}

We begin by providing a brief description of the various QP techniques in  Section~\ref{TheoreticalMethods}.  The specific computational parameters employed for bulk, surface, and interface systems are then listed in Section~\ref{ComputationalDetails}.  Finally, in Section~\ref{ExperimentalEnergyReferences} we describe how energy references are obtained from the experimental data.

\subsection{Theoretical Methods}\label{TheoreticalMethods}

The QP $G_0W_0$ approach involves the single-shot correction of the DFT eigenvalues by the self energy $\Sigma = i G W$, where $G$ is the Green's function and $W$ is the screening\cite{GW}. $W$ is obtained from  the  dielectric function, based on the Kohn-Sham wavefunctions \cite{AngelGWReview}.  This is calculated using linear response time-dependent DFT within the random phase approximation (RPA), including local field effects\cite{KresseG0W0}.  Although one obtains a first-order approximation to the QP eigenvalues within $G_0W_0$, this technique provides no information on how screening affects the energy of the vacuum level, $E_{\mathit{vac}}$, and the spatial distribution of the wavefunctions, i.e.\ the QP wavefunctions.  For image potential levels and wet electron levels the QP wavefunctions are qualitatively different from their Kohn-Sham counterparts \cite{ImageStatesLiFMgO}, motivating the use of self-consistent $GW$ techniques.

In the self-consistent procedure we use \cite{SchilfgaardeQSGW,QSGW,KressescGW}, the QP wavefunctions at each iteration are obtained by diagonalizing the Hermitian part of the Hamiltonian and overlap matrices in the basis of the previous step's wavefunctions.  Here, we denote this methodology by ``scQP$GW$'' as opposed to ``sc$GW$''. This is to provide a clear distinction from self-consistent $GW$ calculations where the full Hamiltonian is diagonalized \cite{CarusoPRB2012}.

At the $\mathrm{scQP}GW$ level, self-consistent $GW$ calculations are performed until full self consistency is obtained for the QP eigenvalues \cite{KresseGW, KressescGW}.  This method has been found to significantly overestimate band gaps for most materials \cite{KressescGW}.  

The $\mathrm{scQP}GW_0$ technique has been proposed to partly remedy $\mathrm{scQP}GW$'s tendency to overestimate band gaps \cite{KressescGW,KresseGW}.  In $\mathrm{scQP}GW_0$ the screening $W$ is fixed to that obtained within RPA based on the Kohn-Sham wavefunctions, i.e. $W_0$.  This is justified by the fact that DFT typically yields reliable dielectric constants \cite{KresseGW}.

In fact, the overestimation of band gaps by self-consistent $GW$ calculations has recently been attributed to the neglect of the lattice polarization contribution to the screening of the electron-electron interaction \cite{MarquesLatticePolarization}.  This effect has been shown to always reduce the band gap, and is particularly significant for polar materials such as TiO$_2$.  Overall, the good performance of $G_0W_0$ is attributable to ``a partial cancellation of errors: the underestimation of the band gap opening is compensated by the neglect of the band gap shrinkage due to the lattice polarization'', as stated in ref.~\citenum{MarquesLatticePolarization}\nocite{MarquesLatticePolarization}.

For this reason, we have recently introduced the $\mathrm{scQP}GW1$ approach \cite{OurJACS}.  The self energy corrections must be applied in small fractions to obtain a smooth convergence of the QP wavefunctions during the self-consistent cycle.  At each step, the eigenvalues and wavefunctions are computed and updated for the subsequent step.  This means one may choose to stop the self-consistent procedure once a full portion of QP self energy has been introduced.  At this point the xc-potential has been completely replaced by self energy.  In this way one obtains QP eigenvalues comparable to those from $G_0W_0$, along with the QP wavefunctions and vacuum level.  If instead the full self energy correction is applied in one step, entirely replacing the xc-potential, one would obtain the same QP eigenvalues as from $G_0W_0$. Here the vacuum level is obtained from the Hartree and ionic electrostatic contributions to the effective potential far from the surface.  In general, $\mathrm{scQP}GW1$ is expected to provide an accurate description of all materials for which $G_0W_0$ calculations have proven successful.

It has previously been shown\cite{AmilcareTiO2GW,HybertsenTiO2GW,OurJACSSI,TiO2GWVasp} that the experimental optical spectra for bulk rutile TiO$_{\text{2}}$ may be obtained via the
Bethe-Salpeter equation (BSE) \cite{KresseBSE} based on $G_0W_0$ eigenvalues. The electrostatic electron-hole interaction is included using an effective nonlocal frequency independent exchange correlation $f_{xc}({\mathbf{r}},{\mathbf{r}}',\omega=0)$ kernel suggested in ref.~\citenum{Reiningfxc}\nocite{Reiningfxc}.   

Similarly, the test charge/test charge $\mathrm{scQP}GW$ scheme, $\mathrm{scQP}GW^{\mathrm{TCTC}}$, includes electrostatic electron-hole interactions, i.e. vertex corrections, within the computation of the screening $W$\cite{KressescGW}. This is again accomplished by including $f_{xc}({\mathbf{r}},{\mathbf{r}}',\omega=0)$ within the dielectric function, as described in ref.~\citenum{Reiningfxc}.  In this work we have applied the $\mathrm{scQP}GW^{\mathrm{TCTC}}$ scheme starting from the converged $\mathrm{scQP}GW$ results.

\subsection{Computational Details}\label{ComputationalDetails}

All calculations have been performed using the DFT code \textsc{vasp}  within the projector augmented wave (PAW) scheme \cite{kresse1999}. We used either a localized density approximation (LDA) \cite{LDA} (bulk), generalized gradient approximation (PBE) \cite{PBE} (surface/interface) or a range separated hybrid functional (HSE) \cite{HSE} (surface/interface) for the xc-functional \cite{kresse1996b}.  In particular, we use the HSE06 variant, with a screening parameter of $\mu = 0.2$~\AA$^{-1}$, of the HSE hybrid xc-functional, which includes 25\% exact-exchange \cite{HSE}.  The geometries have been fully relaxed, with all forces $\lesssim$ 0.02 eV/\AA, a plane-wave energy cutoff of 445 eV, an electronic temperature $k_B T\approx0.2$ eV with all energies extrapolated to $T\rightarrow 0$ K, and a PAW pseudopotential for Ti which includes the 3$s^2$ and 3$p^6$ semi-core levels \cite{HybertsenTiO2GW}.

\subsubsection{Bulk}
The $G_0W_0$ calculations for bulk rutile TiO$_2$ are based on DFT calculations
performed using the LDA \cite{LDA} xc-functional. A $4.5941 \times 4.5941 \times 2.958$~\AA$^3$ unit cell with $D_{2h}$ symmetry, corresponding to the
experimental lattice parameters for bulk rutile TiO$_2$, was employed.  A $\Gamma$ centered \textbf{k}-point mesh of $6\times 6 \times 10$ was used, yielding a sampling of 0.228 \AA$^{-1}$ in the (100)/(010) directions and 0.212 \AA$^{-1}$ in the (001) direction of the Brillouin zone.  The electronic density and wave functions were calculated with an increasing number of unoccupied bands per atom, $n_{\mathrm{unocc}} = 12$, $22\nicefrac{2}{3}$, $38\nicefrac{2}{3}$, $49\nicefrac{1}{3}$, $65\nicefrac{1}{3}$, $76$, $102\nicefrac{2}{3}$, $129\nicefrac{1}{3}$, and $150\nicefrac{2}{3}$, i.e.\ including all levels up to 52, 85, 127, 151, 184, 204, 357, 392, and 428~eV above the VBM, respectively,  to converge the calculation at the $G_0W_0$ level.  We used an energy cutoff for the number of \textbf{G}-vectors for representing the response function of 297 eV, and 192 sampling points for the RPA dielectric function.

As a further comparison, we have performed similar $G_0W_0$ calculations for bulk rutile TiO$_2$ using the Grid-based PAW code \textsc{gpaw} \cite{gpaw1,gpaw2}. Here we used a $7\times7\times11$ \textbf{k}-point mesh, a $h \approx 0.2$~\AA\ grid spacing, an energy cutoff for the number of \textbf{G}-vectors for representing the response function of 100 eV, and 73 sampling points for the RPA dielectric function.

The $\mathrm{scQP}GW1$ calculations for bulk rutile TiO$_2$ are based on DFT calculations performed using the LDA\cite{LDA} and HSE\cite{HSE} xc-functionals including $n_{\mathrm{unocc}} = 12$ unoccupied bands per atom.

For the BSE calculations we used a denser $\Gamma$-centered $8 \times 8 \times 12$ \textbf{k}-point mesh and 480 sampling points for the RPA dielectric function.  We included $n_{\mathrm{unocc}} = 12$ unoccupied bands per atom for the $G_0W_0$ calculation, and included transitions between 12 occupied and 16 unoccupied bands in the BSE calculation \cite{KresseBSE}.

\subsubsection{Surface}
The $G_0W_0$, $\mathrm{scQP}GW1$, $\mathrm{scQP}GW$, and $\mathrm{scQP}GW^{\mathrm{TCTC}}$ calculations for rutile TiO$_2$(110) are based on DFT calculations
performed using the PBE  \cite{PBE} and HSE \cite{HSE} xc-functionals, with $\mathrm{scQP}GW_0$ based on PBE.  We used a four layer pristine TiO$_2$(110) $1\times1$ unit cell of $6.497 \times 2.958  \times 40$ \AA$^3$ with a vacuum separation of  27 \AA. We employed a $\Gamma$-centered $4\times 8\times1$ \textbf{k}-point mesh, 320 bands = 9\nicefrac{1}{3} unoccupied bands per atom, i.e.\ including all levels up to 26~eV above the VBM, an energy cutoff of 80 eV for the number of \textbf{G}-vectors, and a sampling of 80 frequency points for the dielectric function.  

The electronic density and wave functions were also calculated with an increasing number of unoccupied bands per atom, $n_{\mathrm{unocc}} =  22\nicefrac{2}{3}$, $38\nicefrac{2}{3}$, $49\nicefrac{1}{3}$, $65\nicefrac{1}{3}$, and $76$, i.e.\ including all levels up to 46, 66, 77, 93, and 103~eV above the VBM, respectively,  to converge the calculation at the $G_0W_0$ level.

As a further check, we performed PBE $G_0W_0$ calculations for an eight layer pristine TiO$_2$(110) $1\times1$ unit cell of $6.497\times2.968\times53$~\AA$^3$ with a vacuum separation of 27~\AA.  We used either $9\nicefrac{1}{3}$ or $76$ unoccupied bands per atom, i.e. including all levels up to 32 or 134~eV above the VBM.

\subsubsection{Interface}
The $G_0W_0$ calculations  for CH$_3$OH on TiO$_2$(110) are based on DFT calculations
performed using the PBE  \cite{PBE} and HSE \cite{HSE} xc-functionals, with $\mathrm{scQP}GW1$, $\mathrm{scQP}GW$, $\mathrm{scQP}GW_0$ based on PBE. We modelled the most stable monolayer structure of CH$_3$OH on TiO$_{\text{2}}$(110)\cite{Jin} with adsorbates on both sides of a four layer slab, with $C_{2}$ symmetry.  We used a $1\times2$ unit cell of $6.497 \times 5.916 \times 47.0$~\AA$^3$, corresponding to the experimental lattice parameters for bulk rutile TiO$_{\text{2}}$ \cite{TiO2LatticeParameters} in the surface plane.  This provides $\gtrsim 27$~\AA\ of vacuum between repeated images. We employed a $\Gamma$ centered $4\times4\times1$ \textbf{k}-point mesh, with 880 bands = 9\nicefrac{1}{6} unoccupied bands per atom, i.e.\ including all levels up to 30~eV above the VBM, an energy cutoff of 80 eV for the number of \textbf{G}-vectors, and a sampling of 80 frequency points for the dielectric function.

\subsection{Experimental Energy References}\label{ExperimentalEnergyReferences}

Experimental spectra are typically referred to the Fermi level, $\varepsilon_F$, which is pinned about $0.1$~eV below the CBM for mildly reduced TiO$_{\text{2}}$ \cite{TiO2Fermi,TiO2FermiAono,TiO2FermiYamakata}.  Using the experimental work function $\phi \approx 5.3$ -- $5.5$~eV \cite{TiO2WorkFunction30,Onishi198833,C0CP02835E,Petek2PPH2O}, one may obtain the CBM energy relative to the vacuum level of $\varepsilon_{\mathrm{CBM}} \approx -\phi + 0.1 \approx -5.2$ or $-5.4$~eV.  Similarly, using the electronic band gap for rutile TiO$_{\text{2}}$ of $3.3 \pm0.5$~eV obtained from electron spectroscopy measurements \cite{TiO2BandGap}, one may estimate the VBM energy relative to the vacuum level at $\varepsilon_{\mathrm{VBM}} \approx -\phi + 0.1 - 3.3 \approx -8.5$ or $-8.7$~eV. For bare TiO$_\text{2}$(110) we have used these references to compare with the computed VBM and CBM energies relative to the vacuum level.  

For the interface, the workfunction is strongly dependent on the structure of the adsorbing CH$_3$OH ML, i.e.\ $\Delta\phi \sim 0.9$~eV\cite{Jin}.  Moreover, the VBM is the most reliable theoretical reference.  For these reasons, we have aligned the experimental UPS and 2PP spectra for the interface to the VBM.  This is done by adding the value for the Fermi level relative to the VBM, i.e., $\varepsilon_{\mathrm{F}}\approx-0.1+3.3=3.2$~eV.  In this way we are able to directly compare the methanol occupied and unoccupied densities of states we calculate with the UPS and 2PP spectra. 

\section{RESULTS AND DISCUSSION}

For a theoretical technique to be robust for describing interfacial level alignment, it would ideally reproduce the electronic properties of the bulk material and bare surface.  With these ideas in mind, we discuss in Section~\ref{optical} the convergence of the electronic and optical band gap, along with the optical spectra, for bulk rutile TiO$_2$.  Based on these results, we compare the performance of various QP techniques for describing the VBM and CBM level alignment for the bare TiO$_2$(110) surface in Section~\ref{LevelAlignmentBare}.  We then show in Section~\ref{LevelAlignmentMethanol} which of these techniques provide the best agreement with the UPS and 2PP spectra for a CH$_3$OH ML on TiO$_2$(110).  Finally, in Section~\ref{Correlations} we show how the $\mathrm{QP}G_0W_0$ energy corrections are correlated with the spatial distribution of the wave functions.

\subsection{Electronic and Optical Band Gap of Bulk Rutile TiO$_{\text{2}}$}\label{optical}

\begin{figure}[!t]
\includegraphics[width=0.93\columnwidth]{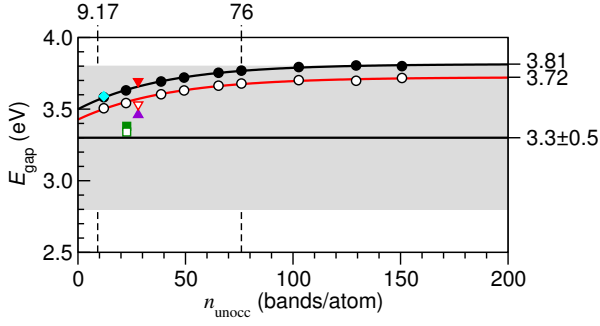}
\caption{
Convergence of the TiO$_2$ bulk rutile $G_0W_0$  direct (filled symbols) and indirect (open symbols) band gaps, $E_{\mathrm{gap}}$, with the number of unoccupied bands per atom, $n_{\mathrm{unocc}}$. 
 Exponential fits (solid lines) yield  asymptotic limits of 3.81 and 3.72~eV for the direct and indirect band gaps, respectively.  Results are from this work with \textsc{vasp} (black circles), \textsc{gpaw} (red triangle down), ref.~\citenum{AmilcareTiO2GW} (cyan diamond), ref.~\citenum{HybertsenTiO2GW} (green squares), and ref.~\citenum{TiO2GWVasp} (violet triangle up). The experimental electronic band gap of $3.3 \pm 0.5$~eV from ref.~\citenum{TiO2BandGap} is provided for comparison.
}\label{BulkConvergence}
\end{figure}
\nocite{TiO2GWVasp}
The first robustness criteria we shall consider is the electronic and optical band gap description for bulk rutile TiO$_2$.  This also includes a comparison between the calculated and measured optical spectra.  

As shown in Figure~\ref{BulkConvergence}, to obtain a reasonable convergence of the direct and indirect band gaps for bulk TiO$_2$ requires about $n_{\mathrm{unocc}} \approx 76$ unoccupied bands per atom.  The convergence with the number of unoccupied bands is quite well described by a simple exponential fit, i.e.\ \(E_\infty + \alpha \exp(-n_{\mathrm{unocc}}/\beta)\). From this we obtain the asymptotic limits of 3.72 eV and 3.81 eV for the direct and indirect gap, respectively.  

\begin{figure}[!t]
\includegraphics[width=\columnwidth]{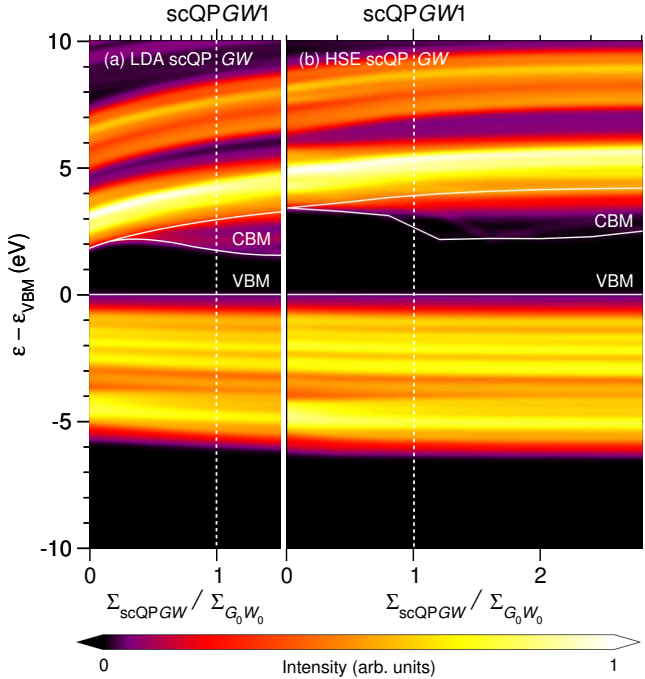}
\caption{Convergence of the DOS for bulk rutile TiO$_{\mathrm{2}}$ with respect to the cumulative sum of portions of self energy introduced self-consistently, $\Sigma_{\mathrm{scQP}GW} / \Sigma_{G_0W_0}$, from (a) LDA $\mathrm{scQP}GW$ and (b)  HSE $\mathrm{scQP}GW$. Energies $\varepsilon$ in eV are taken relative to the VBM $\varepsilon_{\mathrm{VBM}}$. White solid lines indicate the VBM and CBM positions for the direct and indirect gaps. White dashed lines indicate the DOS corresponding to $\mathrm{scQP}GW1$ results.  Upper ticks indicate the steps of the scQP$GW$ calculations.}\label{bulk_conv}
{\color{JPCTCGreen}{\rule{\columnwidth}{1.0pt}}}
\end{figure}

Figure \ref{bulk_conv} shows the convergence of the total density of states (DOS) with respect to the cumulative sum of portions of self energy introduced self-consistently with the QP LDA and HSE $\mathrm{scQP}GW$ calculations.  The upper ticks in Figure~\ref{bulk_conv} denote the steps at which the eigenvalues and QP wavefunctions are calculated.  The intermediate DOS are obtained by linearly interpolating betweeen the calculated eigenvalues at each step.  At \(\Gamma\) the LDA DOS at \(\Sigma_{\mathrm{scQP}GW}/\Sigma_{G_0W_0} = 1.5\) approaches the HSE DFT DOS.  The direct band gap at \(\Sigma_{\mathrm{scQP}GW}/\Sigma_{G_0W_0} = 2.8\) for HSE $\mathrm{scQP}GW$ is $E_{\mathrm{gap}} \approx 4.18$~eV, in agreement with the LDA $\mathrm{scQP}GW$ results of ref.~\citenum{QSGW}. This is related to the starting-point independence of the $\mathrm{scQP}GW$ procedure.  

The indirect band gap is significantly underestimated for both LDA and HSE bulk $\mathrm{scQP}GW$ calculations.  We attribute this to the lack of hermiticity of the  Hamiltonian for these \textbf{k}-points.  This problem is more significant for LDA than HSE $\mathrm{scQP}GW$ calculations.  We suggest this is because HSE DFT is closer to converged $\mathrm{scQP}GW$ than LDA DFT.  Since the CBM levels which make up the indirect band gap are not present in a four layer slab model, these difficulties for the $\mathrm{scQP}GW$ method are not observed for the bare and CH$_3$OH covered TiO$_2$(110) surfaces.

$G_0W_0$ with a converged number of bands significantly overestimates the experimental band gap of $3.3\pm0.5$~eV \cite{TiO2BandGap}, although it is still within the upper limit of the experimental error.  We attribute this overestimation of the electronic band gap to the neglect of lattice polarization contributions within the screening.  This is expected to be quite important for polar materials such as titania.  

The success of $G_0W_0$ for describing the electronic band gaps of polar materials is partly due to a cancellation of errors \cite{Marques}.   Performing $GW$ self-consistently \emph{increases} the $G_0W_0$ electronic gap, while including lattice polarization within the screening significantly \emph{decreases} the electronic gap back to the $G_0W_0$ value.  However, for titania the contribution of lattice polarization within the dielectric function is so significant that the electronic gap is already overestimated at the $G_0W_0$ level.  

Including lattice polarization contributions within the screening requires a self-consistent $GW$ calculation, and is not currently available.  Instead, we have employed a semi-empirical approach.

\begin{figure}[!t]
\includegraphics[width=\columnwidth]{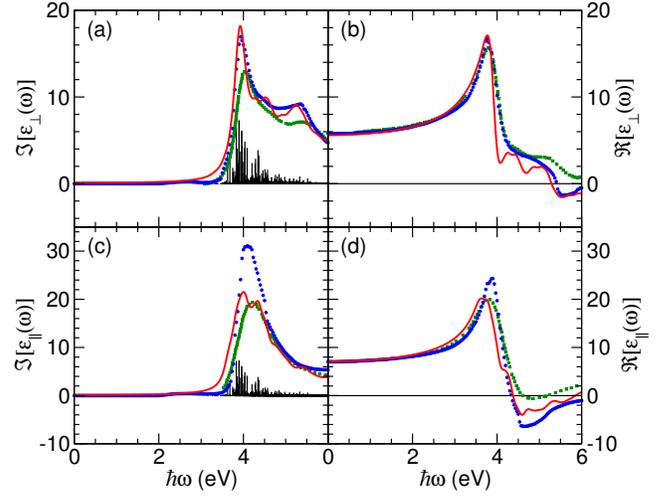}
\caption{Calculated imaginary (a,c) and real (b,d) parts of the dielectric function of bulk rutile TiO$_2$ for polarization perpendicular (a,b) and parallel (c,d) to the TiO$_2$ tetragonal $c$-axis, $\Im[\varepsilon_\perp(\omega)]$\cite{OurJACSSI}, $\Re[\varepsilon_\perp(\omega)]$, $\Im[\varepsilon_\|(\omega)]$, and $\Re[\varepsilon_\|(\omega)]$, versus energy, $\hbar\omega$, in eV. The BSE spectra (red solid line) are based on $G_0W_0$ eigenvalues. The experimental spectra  are obtained from the measured refractive index and absorption index from ref.~\citenum{OpticalTiO2} (blue circles) and ellipsometry measurements of the dielectric function from ref.~\citenum{SPIE} (green squares) at room temperature. 
}
\label{BSE}
{\color{JPCTCGreen}{\rule{\columnwidth}{1.0pt}}}
\end{figure}\nocite{SPIE}

As shown in Figure~\ref{BulkConvergence}, the number of bands employed in previous calculations\cite{AmilcareTiO2GW,HybertsenTiO2GW,TiO2GWVasp} is significantly smaller than that required to fully converge a $G_0W_0$ calculation.   Taking the number of bands to be a ``tuning'' parameter, we have used a reduced number of unoccupied bands within the BSE calculation shown in Figure~\ref{BSE}.

We find 12 unoccupied levels per atom yields a
QP energy gap of 3.32 eV, in quantitative agreement with the experimental band gap of $3.3\pm0.5$~eV \cite{TiO2BandGap}.  
These $G_0W_0$ results have been used in the BSE to calculate the optical absorption spectrum. 

Figure~\ref{BSE} shows the imaginary and real parts of the dielectric function of bulk rutile TiO$_2$ for polarization perpendicular and parallel to the tetragonal axis $c$. This is calculated from the BSE while the experimental results are obtained from optical reflectivity \cite{OpticalTiO2} and ellipsometry \cite{SPIE} measurements at room temperature. The BSE results are in excellent agreement with experiment in both directions: the onsets, the intensities, the main peaks around 4 eV, the line shapes, the macroscopic dielectric function $\Re[\varepsilon(\omega=0)]$, and the plasmon frequencies $\omega_p$ (i.e.\ where $\Re[\varepsilon(\omega_p)] = 0$) are all well reproduced. Further, our $G_0W_0$ and BSE calculations for bulk rutile TiO$_2$ agree with those reported previously\cite{HybertsenTiO2GW,AmilcareTiO2GW,TiO2GWVasp}.

Overall, the success of this semi-empirical approach for the bulk is demonstrated by reproducing the electronic and optical band gaps along with the optical spectrum.  Given the success of this strategy for the bulk, we apply a similarly reduced number of unoccupied bands per atom for our calculations of the surface and interface.  Specifically, we perform QP calculations with $n_{\mathrm{unocc}}=9\nicefrac{1}{3}$ and $9\nicefrac{1}{6}$ per atom for the bare and CH$_3$OH covered TiO$_2$(110) surfaces, respectively.  In this way, we expect to obtain an improved description of the titania level alignment at a significantly reduced computational cost.

\subsection{Level Alignment for Bare Rutile TiO$_{\textbf{2}}$(110)}\label{LevelAlignmentBare}

\begin{figure}[!t]
\includegraphics[width=\columnwidth]{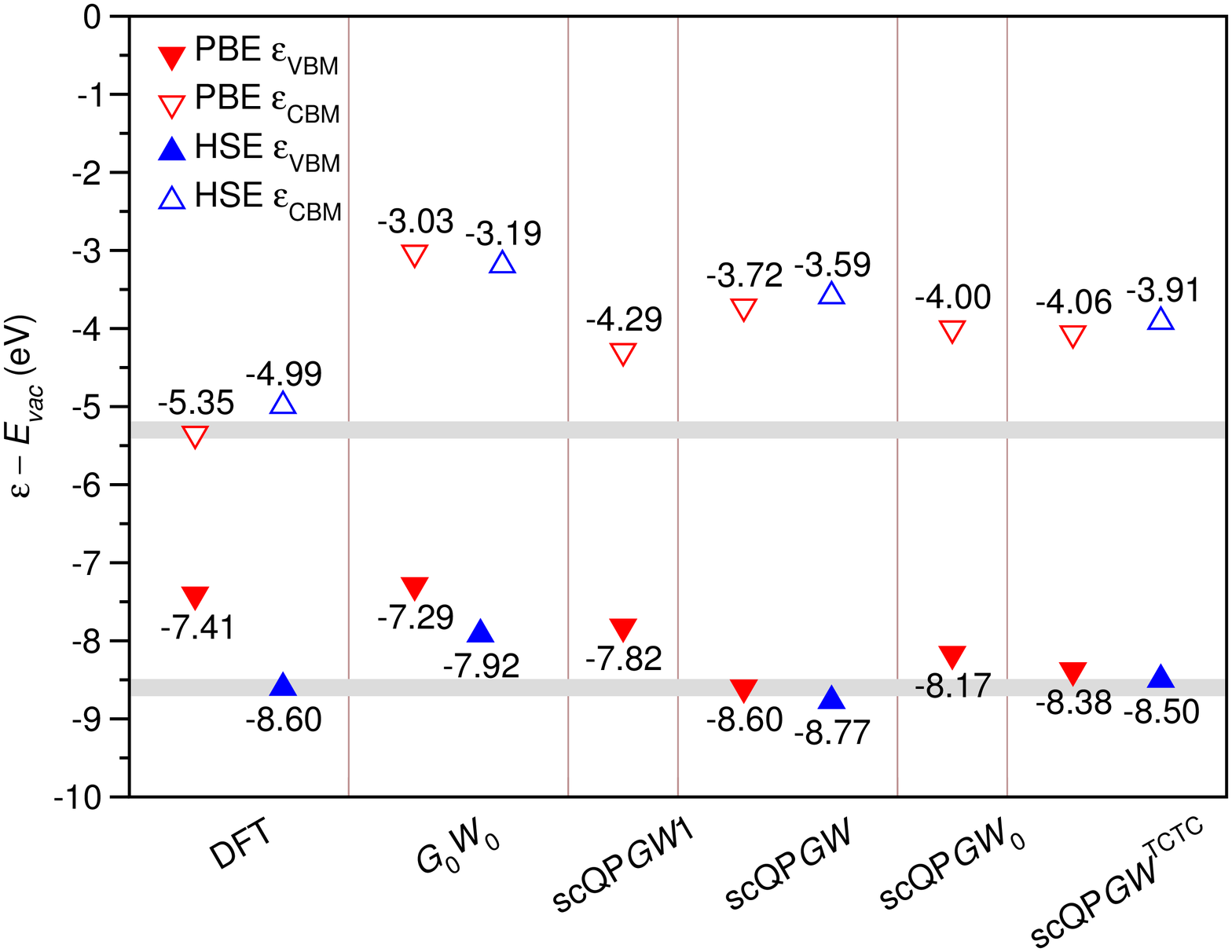}
\caption{VBM and CBM energies, $\varepsilon_{\mathrm{VBM}}$ and $\varepsilon_{\mathrm{CBM}}$, in eV for a bare TiO$_{2}$(110) surface  relative to the vacuum level $E_{\mathit{vac}}$ from DFT, $G_0W_0$, $\mathrm{scQP}GW1$, $\mathrm{scQP}GW$, $\mathrm{scQP}GW_0$, and $\mathrm{scQP}GW^{\mathrm{TCTC}}$ using PBE and HSE xc-functionals.  Gray regions denote VBM and CBM energies derived from the experimental results as discussed in Section \ref{ExperimentalEnergyReferences}.}\label{bare_compare}
{\color{JPCTCGreen}{\rule{\columnwidth}{1.0pt}}}
\end{figure}

The second robustness criteria we shall consider is the alignment of the VBM and CBM for the bare TiO$_2$(110) surface.  The VBM and CBM alignment from DFT, $G_0W_0$, $\mathrm{scQP}GW1$, $\mathrm{scQP}GW$, $\mathrm{scQP}GW_0$, and $\mathrm{scQP}GW^{\mathrm{TCTC}}$ calculations using PBE and HSE xc-functionals is shown in Figure~\ref{bare_compare}. These results are compared with the experimental references for the VBM and CBM discussed in Section~\ref{ExperimentalEnergyReferences}.   

As shown in Figure~\ref{bare_compare}, the computed electronic band gap for bulk rutile TiO$_2$ is significantly smaller than that obtained from a TiO$_2$(110) four layer slab model.  Overall, we find all self-consistent QP techniques describe the VBM level alignment consistently with experiment.  This agrees with the improved description of molecular vertical ionization potentials obtained with self-consistent $GW$ \cite{CarusoPRB2012}.  The band gap overestimation by self-consistent QP techniques is mostly reflected in an overestimation of the CBM energy. For  $G_0W_0$ calculations  the vacuum level is not accessible, so the alignment is relative to the DFT vacuum level. This lack of a well defined vacuum level negatively impacts the absolute energy alignment for $G_0W_0$. Calculating directly the QP vacuum level is a distinct advantage of self-consistent QP techniques.

\begin{figure}[!t]
\includegraphics[width=\columnwidth]{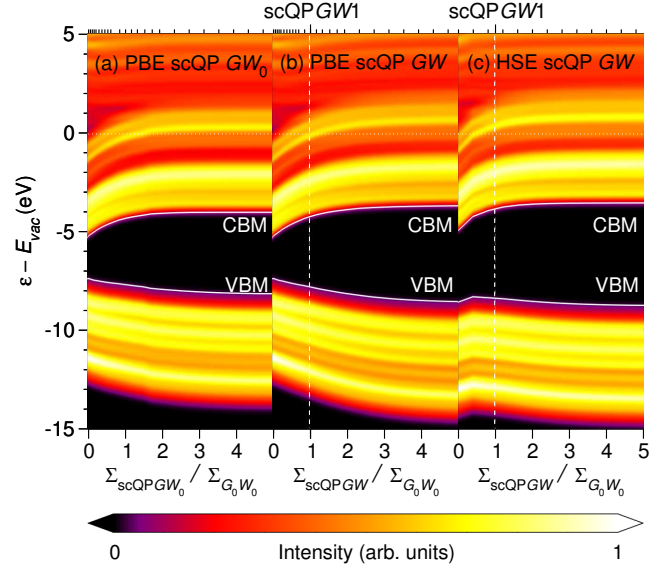}
\caption{Convergence of the DOS for bare TiO$_{\mathrm{2}}$(110) with respect to the cumulative sum of portions of self energy introduced self-consistently, $\Sigma_{\mathrm{scQP}GW} / \Sigma_{G_0W_0}$, from (a) PBE $\mathrm{scQP}GW_0$, (b) PBE $\mathrm{scQP}GW$, and (c) HSE $\mathrm{scQP}GW$. Energies $\varepsilon$ in eV are taken relative to the vacuum level $E_{\mathit{vac}}$ (white dotted line). White solid lines indicate the VBM and CBM positions.  White dashed lines indicate the DOS corresponding to $\mathrm{scQP}GW1$ results.  Upper ticks indicate the steps of the (a) scQP$GW_0$ and (b,c) scQP$GW$ calculations.}\label{surf_conv}
{\color{JPCTCGreen}{\rule{\columnwidth}{1.0pt}}}
\end{figure}

DFT based on the HSE xc-functional (HSE DFT) reproduces the VBM and CBM to within 0.2~eV. The VBM of TiO$_2$(110) is composed of interior O 2$p$ levels, while the CBM is composed of interior Ti $3d$ levels.  As both the VBM and CBM are located within the bulk, a correct energy alignment of these levels requires an accurate description of electron-electron correlation within the bulk.  HSE correlation is basically a constant screening, i.e. an optical dielectric constant of \(\varepsilon_\infty \sim 4\), of the Hartree-Fock exact-exchange term \cite{Marques}.  This provides an improved description for bulk rutile TiO$_2$, for which the measured optical dielectric constants are $\varepsilon_{\infty,a} = 6.84$ and $\varepsilon_{\infty,c} = 8.43$ along the $a$-axis and $c$-axis, respectively\cite{TiO2DielectricConstantExp}. For this reason, HSE works well for the VBM and CBM energy alignment.  Note the RPA optical dielectric constants ($\varepsilon_{\infty,a} \approx 7.83$ and $\varepsilon_{\infty,c} \approx 9.38$)\cite{TiO2DielectricConstantLDA} provide a better description than HSE of the anisotropic screening in bulk rutile TiO$_2$.  Since these anisotropies in the screening are still rather small for the bulk, HSE works reasonably well for the bare TiO$_2$(110) surface.  However, for the interfacial levels of a CH$_3$OH ML on TiO$_2$(110), the anisotropies in the screening are quite important, as we will discuss in Section~\ref{LevelAlignmentMethanol}.

Figure~\ref{surf_conv} shows the convergence of the DOS with respect to the cumulative sum of portions of self energy introduced self-consistently with the QP PBE $\mathrm{scQP}GW_0$, PBE $\mathrm{scQP}GW$, and HSE $\mathrm{scQP}GW$ calculations. From the DFT results, shown to the left of each panel, as the DFT xc-potential is replaced by the self energy $\Sigma$, the gap between occupied and unoccupied levels is increased monotonically.  Most of the correction to the electronic band gap is already introduced at the $\mathrm{scQP}GW1$ level, i.e. when a total of one full ``portion'' of self energy has been included within the self-consistent cycles.  Note that the oscillations observed for HSE $\mathrm{scQP}GW$ are related to the use of a larger step size when introducing the portions of self energy.

The QP energy corrections for the occupied levels mimic those of the VBM. Similarly the corrections for the unoccupied levels with weight in the bulk mimic those of the CBM.  Altogether, this suggests that in the absence of defects, the screening of bulk levels is rather homogeneous for the bare surface.  This means a constant dielectric screening of the Hartree-Fock exact-exchange term may be sufficient to describe the electron-electron correlation for the surface. This is exactly what is accomplished by hybrid xc-functionals such as HSE. This explains the near quantitative agreement for the level alignment with HSE DFT, shown in Figure~\ref{bare_compare}.

However, unoccupied levels with a significant weight in the vacuum, i.e.\ vacuum levels, undergo quite small QP energy corrections.  This is clearly seen from the flat profiles at $\varepsilon \gtrsim 1$~eV above $E_{\mathit{vac}}$.  This is because there is no electronic screening for vacuum levels, so that the bare Hartree term of standard DFT already describes the electron-electron correlation quite well.  

Comparing the PBE $\mathrm{scQP}GW_0$ and $\mathrm{scQP}GW$ results of Figure~\ref{surf_conv} (a) and (b), we see that by fixing the screening to that obtained from DFT, we reach self-consistency earlier, and the opening of the band gap is significantly reduced. Moreover, Figure~\ref{surf_conv} (b) and (c) clearly show the fully converged $\mathrm{scQP}GW$ DOS is independent of the xc-functional (PBE or HSE) used at the DFT level.  Using HSE at the DFT level one reaches self-consistency much earlier, and using larger steps than with PBE.  Altogether, HSE and HSE $G_0W_0$ are closer to $\mathrm{scQP}GW$ than PBE and PBE $G_0W_0$.

\begin{figure}[!t]
\includegraphics[width=\columnwidth]{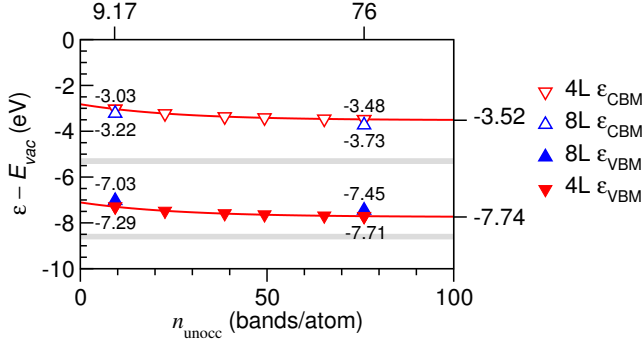}
\caption{Convergence of the VBM and CBM energies, $\varepsilon_{\mathrm{VBM}}$ and $\varepsilon_{\mathrm{CBM}}$, in eV for a bare four layer (4L) and eight layer (8L) TiO$_{2}$(110) surface  relative to the vacuum level $E_{\mathit{vac}}$ from $G_0W_0$, with the number of unoccupied bands per atom, $n_{\mathrm{unocc}}$.  Exponential fits (solid lines) yield asymptotic limits of $-7.74$ and $-3.52$~eV for the VBM and CBM, respectively. Gray regions denote VBM and CBM energies derived from the experimental results as discussed in Section \ref{ExperimentalEnergyReferences}.}\label{SurfaceConvergence}
{\color{JPCTCGreen}{\rule{\columnwidth}{1.0pt}}}
\end{figure}

As a further check, we show in Figure~\ref{SurfaceConvergence} the convergence with number of unoccupied bands of the PBE $G_0W_0$ VBM and CBM energies relative to the DFT vacuum level.  We find the ionization potentials and electron affinities for the bare TiO$_2$(110) surface increase ($\sim 0.4$~eV) with the number of unoccupied bands included in the $G_0W_0$ calculation.  However, the calculated band gap for the four and eight layer surfaces are already converged for $n_{\mathrm{unocc}} = 9\nicefrac{1}{3}$. For four layers with $n_{\mathrm{unocc}} = 9\nicefrac{1}{3}$, $E_{\mathrm{gap}} \approx 4.26$~eV compared to $4.22$~eV in the asymptotic limit.  When the number of layers is increased from four to eight, we obtain the bulk band gap for $n_{\mathrm{unocc}} = 76$ of  3.72~eV.  Although the eight layer slab model provides an improved description of the bulk TiO$_2$ band gap, a four layer model should be sufficient to describe the interfacial level alignment.  This is because the interfacial levels are located primarily within the first few layers of the surface \cite{OurJACS,Migani2014,OurJACSSI}.

The number of bands required to converge the band gap for the bare TiO$_2$(110) surface is much smaller than that needed for bulk rutile TiO$_2$.  This may be due to the different nature of the unoccupied levels being added in $G_0W_0$ calculations for the bulk (e.g.\ Rydberg levels) and the bare surface (e.g.\ vacuum levels).

Overall, we find that the VBM of the bare TiO$_2$(110) surface is described in a consistent manner with experiment by most of the QP techniques considered herein. This is important because the workfunction for methanol covered surfaces is strongly dependent on the structure of the interface. Moreover, the VBM is the most reliable reference from a theoretical perspective.  For these reasons, in Section~\ref{LevelAlignmentMethanol} we carry out the level alignment relative to the VBM. Furthermore, this has the advantage of obtaining convergence of the electronic structure with as few as nine unoccupied bands per atom.

\subsection{Level Alignment for a CH$_{\text{3}}$OH ML on TiO$_{\text{2}}$(110)}\label{LevelAlignmentMethanol}

\begin{figure}[!t]
\includegraphics[width=\columnwidth]{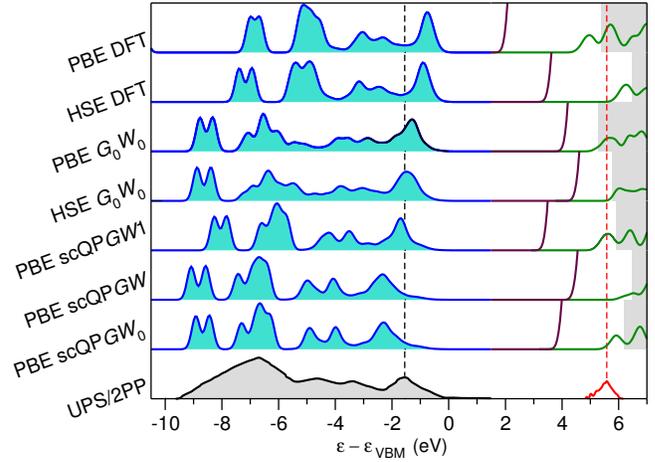}
\caption{Total unoccupied (maroon), CH$_3$OH projected (blue), and Wet (green) DOS  computed with PBE DFT \cite{OurJACSSI}, HSE DFT \cite{OurJACSSI}, PBE $G_0W_0$ \cite{OurJACS}, HSE $G_0W_0$, PBE $\mathrm{scQP}GW1$ \cite{OurJACS}, PBE $\mathrm{scQP}GW$ \cite{OurJACSSI}, PBE $\mathrm{scQP}GW_0$\cite{OurJACS} for an intact methanol monolayer on TiO$_2$(110) and the experimental UPS \cite{Onishi198833}  (black) and 2PP spectra \cite{Onda200532} (red). Filling denotes occupied levels. Energies are relative to the VBM, $\varepsilon_{\mathrm{VBM}}$. Gray shaded regions denote levels above the vacuum level $E_{vac}$.  Black/red dashed vertical lines denote the UPS/2PP highest and lowest energy peaks, respectively.
}\label{int_DOS}
{\color{JPCTCGreen}{\rule{\columnwidth}{1.0pt}}}
\end{figure}

In Sections \ref{optical} and \ref{LevelAlignmentBare} we discussed  the performance of the various QP techniques for describing level alignment with homogeneous screening, i.e., pristine bulk rutile TiO$_2$ and the bare TiO$_2$(110) surface. We now consider the performance of the same QP techniques for the alignment of interfacial levels undergoing anisotropic screening, i.e., the CH$_3$OH ML on TiO$_2$(110) interface shown in Figure~\ref{schematic}. Specifically, in Figure~\ref{int_DOS}, we compare the CH$_3$OH PDOS and Wet DOS computed with PBE DFT, HSE DFT, PBE $G_0W_0$, HSE $G_0W_0$, PBE $\mathrm{scQP}GW1$, PBE $\mathrm{scQP}GW$, and PBE $\mathrm{scQP}GW_0$, with UPS and 2PP spectra for a CH$_3$OH ML on TiO$_2$(110).  

\begin{figure*}
\includegraphics[width=2\columnwidth]{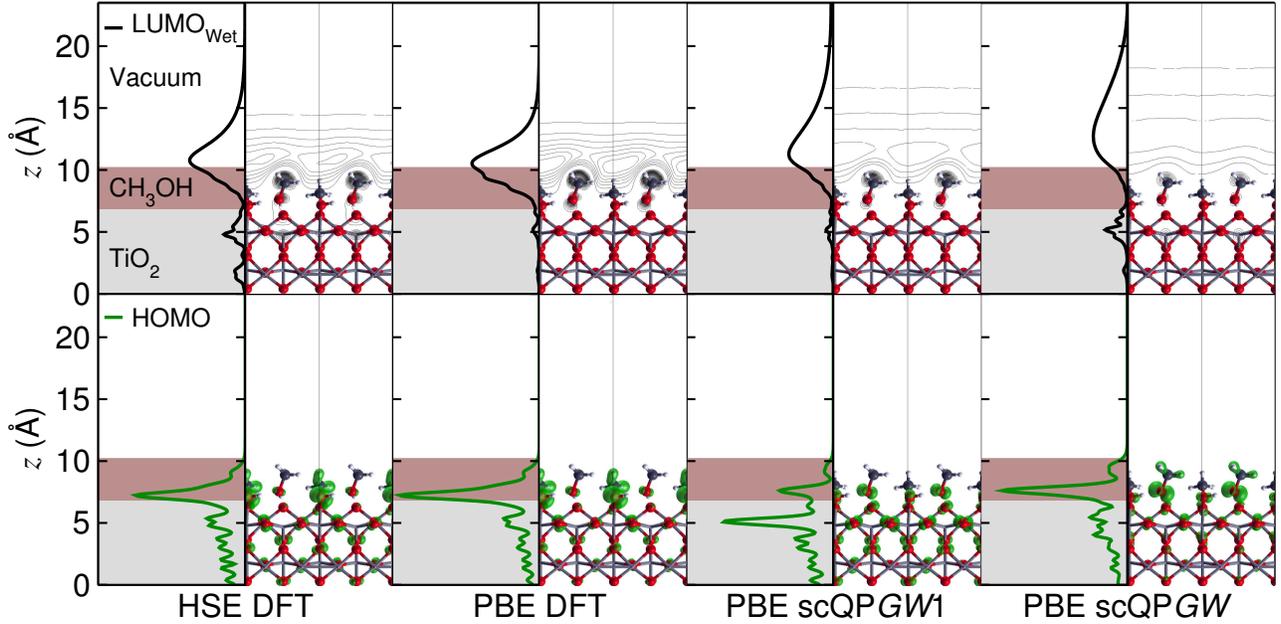}
\caption{
Methanol monolayer on TiO$_{\text{2}}$(110) HOMO (green isosurfaces) and LUMO$_{\textrm{Wet}}$ (black contour plots) average density in the $x y$-plane\cite{OurJACS}, versus distance $z$ in \AA\ from the center of the TiO$_{\text{2}}$ substrate at $\Gamma$ as obtained from HSE DFT \cite{LUMOWet}, PBE DFT \cite{HOMOLUMOWet}, PBE $\mathrm{scQP}GW1$ \cite{HOMOLUMOWet}, and PBE $\mathrm{scQP}GW$ \cite{LUMOWet}. TiO$_{\text{2}}$ bulk, CH$_{\text{3}}$OH molecular layer, and vacuum are depicted by grey, brown, and white regions, respectively.
H, C, O, and Ti atoms are represented by white, grey, red, and silver balls, respectively. 
}\label{wavefunctions}
{\color{JPCTCGreen}{\rule{\textwidth}{1.0pt}}}
\end{figure*}

 UPS experiments\cite{Onishi198833} have probed the occupied molecular levels for the CH$_3$OH ML on TiO$_2$(110) interface. As shown in ref.~\citenum{OurJACS} and depicted in Figure~\ref{wavefunctions}, the higher energy occupied molecular levels (HOMOs) are composed of non-bonding O~2$p$ orbitals of the CH$_3$OH molecules, with some C--H~$\sigma$ and Ti~3$d$ character. 
We have previously shown \cite{OurJACS} that the highest energy peak in the UPS, at $\varepsilon_{\mathit{peak}}^{\mathrm{UPS}}\approx -1.55$~eV relative to the VBM, is due to these  HOMO levels.

Likewise 2PP experiments \cite{PetekScienceMethanol,MethanolSplitting2010,PetekScienceH2O} have probed the unoccupied molecular levels for the CH$_3$OH ML on TiO$_2$(110) interface. As shown in ref.\citenum{OurJACS} and depicted in Figure~\ref{wavefunctions}, these unoccupied molecular levels have a primarily two dimensional (2D) $\sigma^*$ character associated with the methanol C--H bond, with weight above the H atoms outside the molecular layer (brown regions) \cite{PetekScienceMethanol}. 
We have previously shown\cite{OurJACS} that it is these ``Wet electron'' levels \cite{PetekScienceH2O} of intact (i.e., undissociated) CH$_3$OH MLs on TiO$_2$(110), which give the intense experimental peak, at $\varepsilon_{\mathit{peak}}^{\mathrm{2PP}}\approx 5.58$~eV relative to the VBM, in the 2PP spectrum \cite{Onda200532} shown in Figure~\ref{int_DOS}.

As shown in Figure~\ref{wavefunctions}, the Wet electron levels are delocalized within the molecular plane with weight inside and above the molecular layer.  This means they may be screened based on their density averaged over the $x y$-plane.  Specifically, they may be identified as the unoccupied levels with more than half their weight between the bridging O atom of the surface and $5$~\AA\ above the top of the molecular layer. It is these levels which are included in the Wet DOS of Figure~\ref{int_DOS}.

On the one hand, DFT underbinds the HOMO levels, which give the highest energy peak in UPS. In particular, we find the PBE HOMO levels are  closer to the VBM ($\Delta \varepsilon^{\mathrm{UPS}} \approx +0.79$~eV) than those of HSE ($\Delta \varepsilon^{\mathrm{UPS}} \approx +0.65$~eV). On the other hand, PBE overbinds the unoccupied Wet levels  ($\Delta \varepsilon^{\mathrm{2PP}} \approx -0.62$~eV) while HSE underbinds them ($\Delta \varepsilon^{\mathrm{2PP}} \approx +0.79$~eV).

Just as PBE underestimates the electronic band gap of the substrate, it also underestimates the  CH$_3$OH HOMO--LUMO energy gap, although to a much smaller extent. HSE provides an excellent description of the electronic properties of the substrate as described in Section \ref{LevelAlignmentBare}. However, it
applies the same screening to all the levels regardless of their nature. More precisely, HSE applies a homogeneous screening, via the macroscopic dielectric constant \cite{Marques} ($\epsilon_\infty \sim 4$), throughout the unit cell. This fails to describe the anisotropic screening felt by the molecular levels at the interface.  As a result, the unoccupied molecular Wet levels are underbound by  HSE. This means HSE is an inappropriate method for molecular/semiconductor interfaces. Instead, QP techniques, where anisotropic screening is calculated directly, should provide a better description of interfacial levels.

The PBE DFT results shown in Figure~\ref{int_DOS} differ qualitatively from PBE $G_0W_0$.  $G_0W_0$ shifts the PBE energies of the empty levels up and the occupied molecular levels down, giving near-quantitative agreement with the 2PP and UPS results ($\Delta \varepsilon^{\mathrm{UPS}} \approx +0.26$~eV, $\Delta \varepsilon^{\mathrm{2PP}} \approx +0.10$~eV).  As will be shown in Section~\ref{Correlations}, this improved alignment for the interfacial molecular levels is due to a proper description of the anisotropic screening at the interface.

Although at the $G_0W_0$ level the vacuum level and wavefunctions are not directly available, the character of the actual and QP wavefunctions is reflected in the experimental spectra and can be inferred from the calculated $G_0W_0$ PDOS.  For example, the HOMO peak in the $G_0W_0$ PDOS is broadened with respect to the PBE one, in better agreement with experiment (Figure~\ref{int_DOS}).  This indicates that the HOMOs of methanol and the O~2$p_\pi$ levels\cite{DuncanTiO2} of the substrate are more strongly hybridized at the QP level.

The issues with the HSE interfacial level alignment are partially addressed by HSE $G_0W_0$.  
$G_0W_0$ shifts the HSE energies of both the unoccupied and occupied molecular levels down relative to the VBM, giving a better agreement with the 2PP and UPS results ($\Delta\varepsilon^{\mathrm{UPS}} \approx +0.05$~eV, $\Delta\varepsilon^{\mathrm{2PP}}\approx+0.46$~eV). In fact, $G_0W_0$ shifts the HSE VBM up in energy, as shown for the bare TiO$_2$(110) surface in Figures \ref{bare_compare} and \ref{surf_conv}.  This is because the QP shifts for bulk, molecular, and vacuum levels are qualitatively different.  We will provide a more detailed description in Section~\ref{Correlations} of the physical origin and nature of the $G_0W_0$ QP shifts.

To maintain the accurate PBE $G_0W_0$ description of the spectra while also describing the vacuum level and QP wavefunctions via the self-consistent $GW$ procedure, we recently introduced the $\mathrm{scQP}GW1$ approach\cite{OurJACS}. The PBE $\mathrm{scQP}GW1$ spectra shown in Figure~\ref{int_DOS} agree even better than PBE $G_0W_0$ with the UPS and 2PP measurements  ($\Delta\varepsilon^{\mathrm{UPS}} \approx -0.15$~eV, $\Delta\varepsilon^{\mathrm{2PP}}\approx+0.04$~eV). This suggests that the $\mathrm{scQP}GW1$ QP wavefunction may be more representative of the actual wavefunction.

As shown in Figure~\ref{int_DOS}, $\mathrm{scQP}GW$ significantly underbinds the CBM and Wet electron level energies ($\Delta \varepsilon^{\mathrm{2PP}}\approx+0.95$~eV), and overbinds the occupied molecular level energies ($\Delta \varepsilon^{\mathrm{UPS}}\approx-0.78$~eV). Better agreement is obtained from $\mathrm{scQP}GW_0$, which improves the description of the unoccupied levels ($\Delta \varepsilon^{\mathrm{2PP}}\approx+0.33$~eV), but not the occupied molecular levels ($\Delta\varepsilon^{\mathrm{UPS}}\approx-0.75$~eV).  Although $\mathrm{scQP}GW_0$ reduces the $\mathrm{scQP}GW$ band gap by $\sim 0.6$~eV, the occupied levels have similar energies to $\mathrm{scQP}GW$.

\begin{figure}[!t]
\includegraphics[width=0.995\columnwidth]{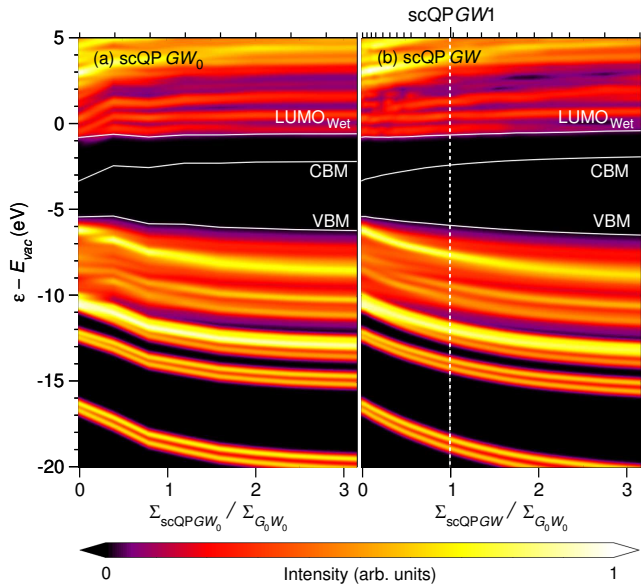}
\caption{Convergence of the PDOS and Wet DOS for CH$_{\mathrm{3}}$OH on TiO$_{\mathrm{2}}$(110) with respect to the cumulative sum of portions of self energy introduced self-consistently, $\Sigma_{\mathrm{scQP}GW} / \Sigma_{G_0W_0}$, from (a) PBE $\mathrm{scQP}GW_0$ and (b) PBE $\mathrm{scQP}GW$. Energies $\varepsilon$ in eV are taken relative to the vacuum level $E_{\mathit{vac}}$. White solid lines indicate the VBM, CBM, and LUMO$_{\mathrm{Wet}}$ positions.  White dashed lines indicate the DOS corresponding to $\mathrm{scQP}GW1$ results.  Upper ticks indicate the steps of the (a) scQP$GW_0$ and (b) scQP$GW$ calculations.}\label{int_conv}
{\color{JPCTCGreen}{\rule{\columnwidth}{1.0pt}}}
\end{figure}

Figure~\ref{int_conv} shows the convergence of the  CH$_3$OH PDOS and Wet DOS with respect to the cumulative sum of portions of self energy introduced self-consistently with the QP PBE $\mathrm{scQP}GW_0$ and PBE $\mathrm{scQP}GW$ calculations.
As for the bare TiO$_2$(110) surface, most of the correction to the electronic band gap is already introduced at the $\mathrm{scQP}GW1$ level, i.e. when a total of one full ``portion'' of self energy has been included within the self-consistent cycles.  $\mathrm{scQP}GW_0$ provides a reduced band gap and  upshifts the occupied CH$_3$OH PDOS and downshifts the WET DOS relative to $\mathrm{scQP}GW$.
Note that the oscillations observed for scQP$GW_0$ are related to the use of a larger step size when introducing the portions of self energy.

The QP corrections of the occupied molecular levels follow those of the HOMO levels, with a steeper descent than the VBM. This is qualitatively different from the occupied bulk levels, which follow the VBM. The same behaviour was shown in Figure~\ref{surf_conv} for the occupied bulk levels of the bare TiO$_2$(110) surface.     

The QP corrections of the Wet electron levels follow the LUMO$_{\mathrm{Wet}}$.  These corrections are qualitatively different from those of the CBM and have a much ``flatter'' energy profile. In particular the LUMO$_{\mathrm{Wet}}$ energy profile's slope is in between those of the CBM and vacuum levels. These differences in slope are a direct consequence of the anisotropic screening at the interface.  In Section~\ref{Correlations} we will show the $G_0W_0$ QP corrections are directly correlated with the spacial distribution of the wavefunction.

Overall, we find PBE $G_0 W_0$ and PBE $\mathrm{scQP}GW1$ provide the correct level alignment for methanol on TiO$_{\text{2}}$(110), while PBE, HSE, HSE $G_0W_0$, PBE $\mathrm{scQP}GW$, and PBE $\mathrm{scQP}GW_0$ deviate from the UPS or 2PP spectra.  At a qualitative level, we find the HOMO energies provide an instructive  ordering with decreasing energy of the QP methods: PBE $>$ HSE $>$ PBE $G_0W_0$ $>$ HSE $G_0W_0$ $>$ PBE $\mathrm{scQP}GW1$ $>$ PBE $\mathrm{scQP}GW_0$ $>$ PBE $\mathrm{scQP}GW$.  The Wet electron energies provide a similar ordering with increasing energy for PBE based QP methods: PBE $<$ PBE $G_0W_0$ $<$ PBE $\mathrm{scQP}GW1$ $<$ PBE $\mathrm{scQP}GW_0$ $<$ HSE $G_0W_0$ $\lesssim$ HSE DFT $\lesssim$ PBE $\mathrm{scQP}GW$. 
 As HSE provides the same homogeneous screening to the Wet electron levels and the unoccupied bulk levels, the HSE DFT and HSE $G_0W_0$ Wet levels are already quite close to the PBE $\mathrm{scQP}GW$ results.  This reflects both the starting point independence of $\mathrm{scQP}GW$ and that HSE provides an electronic structure which is closer to $\mathrm{scQP}GW$ than that from PBE.

Next we show how this ordering of the QP techniques is reflected in the spatial distribution of the HOMO wavefunctions shown in Figure~\ref{wavefunctions}. 
The particular HOMO level we consider is the highest energy orbital with $\gtrsim 30\%$ of its weight on the atomic orbitals of the molecule.
As shown in Figure~\ref{wavefunctions}, for HSE and PBE DFT this HOMO level is  mostly localized on the methanol molecule closer to the surface, i.e., atop the Ti coordinately unsaturated site (cus)\cite{OurJACS}.  For $\mathrm{scQP}GW1$, the HOMO becomes hybridized with the three-fold coordinated oxygen atoms at the surface and the other methanol molecule.  For $\mathrm{scQP}GW$, the weight of the HOMO level is shifted to the other methanol molecule.  This reflects both a reordering of the levels, and an increase in hybridization with the bulk levels from the self-consistent QP techniques.

In Figure~\ref{wavefunctions} we also consider how the LUMO$_{\textrm{Wet}}$ wavefunction changes with the level of theory employed.  In general, we find the LUMO$_{\textrm{Wet}}$ level becomes increasingly delocalized into the vacuum as one successively moves from PBE to HSE, $\mathrm{scQP}GW1$, and finally $\mathrm{scQP}GW$.  
By comparing Figures \ref{int_DOS} and \ref{wavefunctions}, a correlation is clearly evident between the weight of the LUMO$_{\mathrm{Wet}}$ in the vacuum and its energy.  Namely, a greater weight of the wavefunction in the vacuum corresponds to a higher peak energy for the Wet DOS.  This may be understood as a gradual alignment of the Wet DOS with the vacuum level as the levels become increasingly vacuum like.  

At the same time the LUMO$_{\mathrm{Wet}}$ wavefunction also becomes increasingly hybridized with the bulk, with HSE being most similar to $\mathrm{scQP}GW$ in this respect.  
Overall, the LUMO$_{\textrm{Wet}}$ levels change from being molecular levels with $\sigma^*$ character, to increasingly image potential-like vacuum levels at the QP level, as found previously for insulator surfaces \cite{ImageStatesLiFMgO}.

\subsection{Analysis of the QP Corrections for a CH$_{\text{3}}$OH ML on TiO$_{\text{2}}$(110)}\label{Correlations}

\begin{figure}[!t]
\includegraphics[width=\columnwidth]{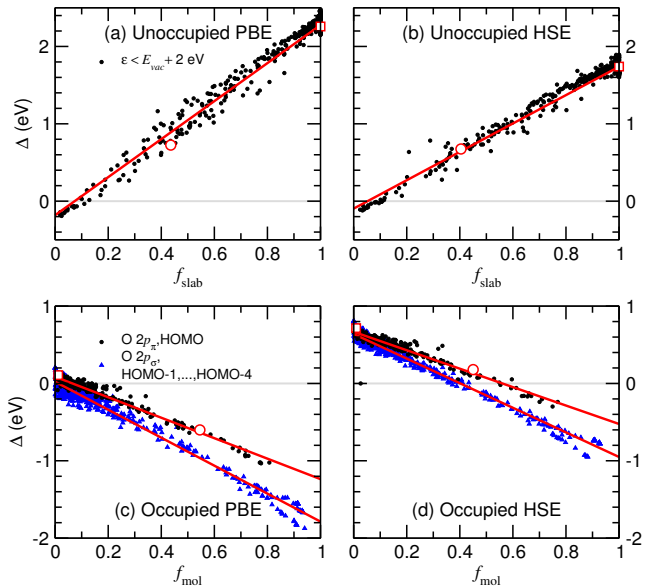}
\caption{
$G_0W_0$ QP energy correction $\Delta$ in eV versus fraction of the wavefunction's density (a,b) in the slab $f_{\mathrm{slab}}$ for the unoccupied levels and (c,d) in the molecular layer $f_{\mathrm{mol}}$ for the occupied levels from (a,c) PBE \cite{OurJACS} and (b,d) HSE for an intact methanol monolayer on TiO$_{2}$(110). Open circles denote the (a,b) LUMO$_{\textrm{Wet}}$ and (c,d) HOMO at $\Gamma$ depicted in Figure~\ref{wavefunctions}.  Open squares denote the (a,b) CBM and (c,d) VBM levels.  Red solid lines are linear fits.  (a,c) Adapted with permission from ref.~\citenum{OurJACS}. Copyright 2013 American Chemical Society.}\label{int_correlations}
{\color{JPCTCGreen}{\rule{\columnwidth}{1.0pt}}}
\end{figure}

In Section \ref{LevelAlignmentMethanol} we showed that a proper description of anisotropic screening is necessary to describe level alignment at an interface.  Following the analysis of ref.~\citenum{OurJACS}, we find the QP $G_0W_0$ energy corrections may be directly related to the spatial distribution of the wavefunctions.  

Figure~\ref{int_correlations} shows that the QP $G_0W_0$ energy shifts for (a) PBE and (b) HSE unoccupied levels are proportional to the fraction of the wavefunction within the bulk or molecular layer, i.e., slab, $f_{\mathrm{slab}}$, (gray and brown regions in Figure~\ref{wavefunctions}).  Similarly, Figure~\ref{int_correlations} shows that the QP $G_0W_0$ energy shifts for (c) PBE and (d) HSE occupied levels are proportional to the fraction of the wavefunction within the molecular layer, $f_{\mathrm{mol}}$, (brown region in Figure~\ref{wavefunctions}).  

For the unoccupied levels, we find for PBE the QP energy shifts are greater for bulk levels than HSE.  This is as expected, since a significant portion of the screening is already included within the HSE xc-functional, resulting in a better CBM energy.  For vacuum levels there is a small QP shift towards stronger binding with PBE, and essentially no QP shift with HSE.  This is because the bare Hartree interaction already describes electron-electron correlation quite well for vacuum levels. Note that the LUMO$_{\mathrm{Wet}}$ level (red open circles in Figure~\ref{int_correlations} (a) and (b)) have nearly equal weights in the slab and vacuum ($f_{\mathrm{slab}} \sim 0.4$). This implies the screening felt by these levels is significantly anisotropic.  This explains why QP $G_0W_0$ is needed to properly describe the 2PP spectra.

As shown in Figure~\ref{int_correlations} (c) and (d), the  occupied levels follow two different correlations  based on their nature. The TiO$_2$ O~$2p$ level are labelled according to ref.~\citenum{DuncanTiO2}\nocite{DuncanTiO2}. The first correlation includes the weaker bound levels, i.e., the VBM, O~2$p_\pi$, and methanol HOMO levels. The second correlation includes more strongly bound levels, i.e.,  the O~2$p_\sigma$, HOMO$-1, \ldots,$ HOMO$-4$ levels.  The correlation is orbital dependent because it is easier to screen $\sigma$ orbitals, which are located between the atoms, than $\pi$ orbitals, which are out of plane \cite{OurJACS}.  

For the occupied bulk levels, we find the QP energy shifts for PBE are almost negligible, while these levels have a significant QP energy shift to weaker binding for HSE.  In particular, the HSE VBM is shifted up by about $0.6$~eV.  This is consistent with what was shown for the bare TiO$_2$(110) surface in Figure~\ref{bare_compare}.  As with the LUMO$_{\mathrm{Wet}}$ level, we find the HOMO level (red open circles in Figure~\ref{int_correlations} (c) and (d)) also have nearly equal weights in the molecular and bulk regions ($f_{\mathrm{mol}} \sim 0.5$).  This implies the screening felt by these levels is also significantly anisotropic.  This explains why QP $G_0W_0$ is needed to properly describe the UPS spectra.  Altogether, these results underline the importance of a proper QP treatment of the anisotropic screening at an interface for the a correct description of interfacial level alignment.

\section{CONCLUSIONS}

We have shown that anisotropic screening is an essential component in determining level alignment of photocatalytically active interfacial systems. A proper treatment of the anisotropic screening necessitates the use of QP techniques. 

For the CH$_3$OH ML on TiO$_2$(110) interface, PBE $\mathrm{scQP}GW1$ is the most accurate technique for reproducing the electronic band gap and the HOMO and LUMO$_\textrm{Wet}$ level alignment. In $\mathrm{scQP}GW1$ the xc-potential is entirely replaced by a full portion of self energy  during the self-consistent QP procedure.  For this reason, the QP energy shifts are quite similar to $G_0W_0$.
 Moreover, $\mathrm{scQP}GW1$ gives one access to the QP wavefunctions and vacuum level, which are lacking in $G_0W_0$. Based on the work of Marques \emph{et al.}, we attribute the excellent performance of PBE $G_0W_0$ and $\mathrm{scQP}GW1$ to a cancellation of opposite contributions.  Specifically, the band gap opening of a fully converged $\mathrm{scQP}GW$ calculation is  counterbalanced by the band gap reduction due to the lattice polarization contribution to the dielectric function \cite{Marques}.  The latter contribution is especially important for polar materials such as TiO$_2$.

On the one hand, the HSE $G_0W_0$ results for the CH$_3$OH ML on TiO$_2$(110) interface are closer to the $\mathrm{scQP}GW$ than PBE $G_0W_0$. This means that we do not have this fortuitous error cancellation for the band gap. On the other hand, HSE DFT works well when the screening felt by the wavefunctions is nearly  homogeneous. We have shown that this is the case for the VBM and CBM level alignment relative to the vacuum for the bare TiO$_2$(110).  In particular, HSE is able to reproduce the MIES work function.  

Finally we have demonstrated that the $G_0W_0$ QP energy corrections for both PBE and HSE are directly related to the spatial distribution of the wavefunction. 
Altogether, these calculations provide a new benchmark for the interpretation of MIES, UPS, and 2PP experiments of complex organic molecule--semiconductor interfaces. 

\section{AUTHOR INFORMATION}
\titleformat{\subsubsection}{\bfseries\sffamily\normalsize}{\thesubsubsection.~}{0pt}{}
\titlespacing{\subsubsection}{0pt}{0pt}{*1}
\subsubsection*{Corresponding Authors}
\noindent *E-mail: annapaola.migani@cin2.es (A.M.).\\ *E-mail: duncan.mowbray@gmail.com (D.J.M.).\\  *E-mail: angel.rubio@ehu.es (A.R.).
\subsubsection*{Notes} 
\noindent The authors declare no competing financial interest.
\section{ACKNOWLEDGEMENTS} 
We acknowledge fruitful discussions with Amilcare Iacomino; funding from the European Projects DYNamo (ERC-2010-AdG-267374) and CRONOS (280879-2 CRONOS CP-FP7); Spanish Grants (FIS2012-37549-C05-02, FIS2010-21282-C02-01, PIB2010US-00652, RYC-2011-09582, JAE DOC, JCI-2010-08156); Grupos Consolidados UPV/EHU del Gobierno Vasco (IT-319-07); NSFC (21003113 and 21121003); MOST (2011CB921404); and NSF Grant CHE-1213189; and computational time from i2basque, BSC Red Espanola de Supercomputacion, and EMSL at PNNL by the DOE.

\bibliography{bibliography}

\end{document}